\documentclass[11pt]{article}
\usepackage[margin=2.5cm,top=1.5cm,bottom=1.5cm]{geometry}

\usepackage{array}
\usepackage{mathtools,multirow}
\usepackage[section]{placeins}
\usepackage{float,caption}
\usepackage{relsize}
\usepackage{makeidx,url,algorithmicx,algorithm,algpseudocode}
\usepackage{amssymb,amsmath,amsthm,color}  
\usepackage{pdfsync,graphicx,subfigure,hyperref}
\usepackage{graphicx}
\usepackage{amsmath}
\usepackage[T1]{fontenc}
\usepackage[english]{babel}
\usepackage[utf8]{inputenc}
\usepackage{apacite}
\usepackage[colorinlistoftodos]{todonotes}
\begin{document}
\title{Scaling priors in two dimensions for Intrinsic Gaussian Markov Random Fields}
\author{
		Maria-Zafeiria Spyropoulou\thanks{Department of Mathematics, Statistics and Actuarial Science, University of Kent, CT2 7FS, UK.},  James Bentham\footnotemark[1]\\
		(Correspondence: {\tt mzs2@kent.ac.uk}.)  
	}
\maketitle
\begin{abstract}	
\noindent Intrinsic Gaussian Markov Random Fields (IGMRFs) can be used to induce conditional dependence in Bayesian hierarchical models. IGMRFs have both a precision matrix, which defines the neighbourhood structure of the model, and a precision, or scaling, parameter. Previous studies have shown the importance of selecting this scaling parameter appropriately for different types of IGMRF, as it can have a substantial impact on posterior results. Here, we focus on the two-dimensional case, where tuning of the parameter is achieved by mapping it to the marginal standard deviation of a two-dimensional IGMRF. We compare the effects of scaling various classes of IGMRF, including an application to blood pressure data using MCMC methods.
\vspace{3mm}
	\end{abstract}

\section{Introduction}
Intrinsic Gaussian Markov Random Fields (IGMRFs) are used widely as prior distributions in Bayesian hierarchical models, particularly for modelling spatial or temporal data, as they capture conditional dependence through their precision matrices \cite{rue2005gaussian}. We examine two-dimensional IGMRFs, which capture dependence between a pair of variables at multiple time points. They are of various types, and can be specified to induce particular neighbourhood structures for the precisions, either by varying weights, introducing certain behaviour at boundaries or within the precision matrix, or by considering different sets of neighbours \cite{terzopoulos1988computation}.   

Our analysis for the two-dimensional case is a generalisation of previous work on scaling different types of field in one dimension, which used an IGMRF as the prior for capturing non-linear trends and a hyperprior for the precision parameter \cite{sorbye2014scaling}. We must select these precision parameters so that the same degree of scaling is applied to bivariate data as in the one-dimensional case. These choices require particular care in two dimensions, where differences in behaviour between IGMRFs may be larger than in a single dimension. We show that appropriate behaviour can be achieved using real data when applying MCMC methodology \cite{spyropoulou2022}. 

The paper is structured as follows. Section 2 describes the behaviour of different IGMRFs, while Section 3 describes the mapping between the precision parameter and the marginal standard deviation for various two-dimensional IGMRFs. An application to blood pressure data is presented in Section 4, with a discussion of our findings and suggestions for future work in Section 5.

\section{Use of IGMRFs as priors}
\subsection{Motivation}
Blood pressure is bivariate, with measurements comprising systolic and diastolic values (SBP and DBP, respectively). While a realistic one-dimensional model of trends at national level has been developed \cite{danaei2011national, finucane2014bayesian}, it cannot estimate SBP and DBP simultaneously, and no information is captured on interactions between the variables. We have developed a two-dimensional extension, including analogous terms to the original model \cite{spyropoulou2022}. Specifically, we have a vector $\textbf{y}_{h,i}$ of SBP and DBP measurements and their interactions, indexed by age group $h$ and study $i$ in country $j$, assumed to be distributed
	\begin{align} \label{norm}
	\textbf{y}_{h,i}&\sim \mathcal{N}(\textbf{a}_{j[i]}+\textbf{b}_{j[i]}t_i+\textbf{u}_{j[i],t_{i}}+\textbf{X}_i\boldsymbol\beta + \boldsymbol\gamma_i(z_h) + \textbf{e}_i, \textbf{SD}_{h,i}^2/n_{h,i} +\boldsymbol{\tau}_{i}^{2})
	\end{align}

The model includes country-level linear intercepts and slopes, $\mathbf{a}_{j[i]}$ and $\mathbf{b}_{j[i]}$, time-varying non-linear terms, $\mathbf{u}_{j[i]}$, covariate effects $\boldsymbol\beta$, terms in age $\boldsymbol\gamma_i$, study-specific random effects $\mathbf{e}_i$, age-varying study-specific random effects $\mathbf{w}_{h,i}$ corresponding to $\boldsymbol\tau_i^2$, and noise $\boldsymbol\epsilon_{h,i}$, assumed iid Gaussian. In the earlier work, a one-dimensional second-order IGMRF was used as a prior for the $\mathbf{u}_{j[i]}$ terms, which we have extended to the two-dimensional case.

\subsection{IGMRFs of one and two dimensions}\label{IGMRFs}
We begin by comparing the behaviour of one-dimensional first-order and second-order IGMRFs with the two-dimensional second-order case \cite{rue2005gaussian}. An IGMRF can be defined as
\begin{equation}
\pi(\textbf{x}) = (2\pi)^{-(n-k)/2}(|\textbf{Q}^{*}|)\exp\left(-\frac{1}{2}(\textbf{x}-\boldsymbol{\mu})^{T}\textbf{Q}(\textbf{x}-\boldsymbol\mu)\right) 
\end{equation}
where $k$ denotes the order of the IGMRF, $n$ is the total number of nodes, and the rank is defined as $n-k$. As described previously \cite{rue2005gaussian}, for a vector of observations $\mathbf{u}$ of length $n$, the one-dimensional first-order model assumes independent first-order increments, and we have
\begin{align} \label{dens}
\Delta u_s = u_{s+1}-u_s\sim \mathcal{N}(0,\lambda^{-1}), \quad s= 1,...,n-1
\end{align}
with joint density
\begin{align}\label{dens2}
\pi (\textbf{u}|\lambda) \propto \lambda^{(n-1)/2}\exp\left(-\dfrac{\lambda}{2}\sum_{s=1}^{n-1}(u_{s+1}-u_{s})^2\right)
\end{align}
The second-order model assumes independent increments
\begin{align}\label{dens3}
\Delta^2 u_s = u_{s+2}-2u_{s+1}+u_s \sim \mathcal{N}(0,\lambda^{-1}),  \quad s= 1,...,n-2
\end{align}
with joint density
\begin{align}\label{dens4}
\pi (\textbf{u}|\lambda) \propto \lambda^{(n-2)/2} \exp \left(-\dfrac{\lambda}{2}\sum_{s=1}^{n-2}(u_{s+2}-2u_{s+1}+u_{s})^2\right)
\end{align}  
In two dimensions, the second-order model constructed on a torus assumes independent two-dimensional second-order increments \cite{rue2005gaussian}, and for variables indexed $d$ and $s$ we have
\begin{align}\label{dens5}
\Delta_0^2u_{d,s} = (\Delta^2_{(1,0)} +\Delta^2_{(0,1)}) u_{d,s} &=
u_{d+2,s}-2u_{d+1,s}+2u_{d,s}-2u_{d,s+1}+u_{d,s+2}
\end{align}
This can be written as
\begin{align}\label{dens61}
\Delta_0^2u_{d,s} = 
u_{d+1,s}-4u_{d,s}+u_{d-1,s}+u_{d,s+1}+u_{d,s-1}
&\sim \mathcal{N}(0,\lambda^{-1})
\end{align}
with joint density
\begin{align}\label{dens6}
\pi (\mathbf{u}|\lambda) \propto \lambda^{(n_1\times n_2-3)/2}\exp\left(-\dfrac{\lambda}{2}\sum_{d=2}^{n_1-1}\sum_{s=2}^{n_2-1}(\Delta^2_{(1,0)}u_{d,s} +\Delta^2_{(0,1)} u_{d,s})^2\right)
\end{align} 
where $n_1$ and $n_2$ represent the total number of nodes for each variable. 

Our models are time-varying, so the assumption of an IGMRF on a torus is not appropriate, and a more suitable two-dimensional second-order density \cite{yue2010nonstationary} is
\begin{equation}\label{dens7}
\begin{split}
\pi (\textbf{u}|\lambda) &\propto \lambda^{(n_1\times n_2-3)/2}\exp\Biggl(-\dfrac{\lambda}{2}
\sum_{d=2}^{n_1-1}\sum_{s=2}^{n_2-1}
	 \lbrace\Delta_0^{2}u_{d,s}\rbrace^2+ \lbrace\Delta_1u_{1,1}\rbrace^2+
	\lbrace\Delta_2u_{n_1,1}\rbrace^2\\
	&+\lbrace\Delta_3u_{1,n_2}\rbrace^2+ \lbrace\Delta_4u_{n_1,n_2}\rbrace^2+
	\sum_{d=2}^{n_1}(\lbrace\Delta_5u_{d,1}\rbrace^2+
	\lbrace\Delta_6u_{d,n_2}\rbrace^2)\\
	& +\sum_{s=2}^{n_2}(\lbrace\Delta_7u_{1,s}\rbrace^2+
	\lbrace\Delta_8u_{n_1,s}\rbrace^2)\Biggr)
\end{split}
\end{equation}
A special case of \eqref{dens7} arises when the variables have the same number of nodes, i.e., $n=n_1=n_2$
\begin{equation}\label{dens8}
\begin{split}
\pi (\textbf{u}|\lambda) &\propto \lambda^{(n^2-3)/2}\exp\Biggl(-\dfrac{\lambda}{2}
	\sum_{d=2}^{n-1}\sum_{s=2}^{n-1} \lbrace\Delta_0^{2}u_{d,s}\rbrace^2+ \lbrace\Delta_1u_{1,1}\rbrace^2+
	\lbrace\Delta_2u_{n,1}\rbrace^2\\
	&+\lbrace\Delta_3u_{1,n}\rbrace^2+ \lbrace\Delta_4u_{n,n}\rbrace^2+
	\sum_{d = 2}^{n}(\lbrace\Delta_5u_{d,1}\rbrace^2+
	\lbrace\Delta_6u_{d,n}\rbrace^2)\\
	& +\sum_{s = 2}^{n}(\lbrace\Delta_7u_{1,s}\rbrace^2+
	\lbrace\Delta_8u_{n,s}\rbrace^2)\Biggr)
\end{split}
\end{equation}
In summary, in each case we have $ \textbf{u} \sim \mathcal{N}(0,(\lambda\textbf{P})^{-1})$, where $\textbf{u}$ represents time-varying non-linear effects. They follow an IGMRF that depends on the structure matrix, $\textbf{P}$, and a precision parameter, $\lambda$, which is a scalar in both the one- and two-dimensional cases \cite{yue2010nonstationary}. 

\section{Specifying hyperpriors for two-dimensional IGMRFs}
Given that the structure matrices and marginal variances of IGMRFs vary depending on their type, hyperpriors need to be chosen and assigned appropriate ranges for a particular model based on its structure, particularly its dimensionality and the number of nodes considered \cite{sorbye2014scaling}. For example, in our model, we need to scale $\lambda$ appropriately for a two-dimensional second-order IGMRF with boundaries, which has up to 40 nodes. Here, we derive the reference standard deviation and use it to select appropriate values for a specific hyperprior. 
\subsection{Reference standard deviation} 
We can describe an IGMRF using an alternative definition
\begin{equation}\label{transf1}
\begin{split}
\mathbf{u} \sim \mathcal{N}\left(\textbf{0}, (\lambda \textbf{P})^{-1}\right),\hspace{3mm}\sigma^{2}_{\lambda}(u_i) = \lambda^{-1} \Sigma^{2*}_{ii}
\end{split}
\end{equation}
where $\textbf{P}$ is the structure matrix of the precision matrix and $\Sigma_{ii}$ is the diagonal element of the covariance matrix in position $i$. For the standardized normal distribution, we have
\begin{equation}\label{transf2}
\begin{split}
\mathbf{u} \sqrt{\lambda} \sim \mathcal{N}\left(\textbf{0}, \textbf{P}^{-1}\right),\hspace{3mm}
\lambda\sigma^{2}_{\lambda}(u_i) = \Sigma^{2*}_{ii} 
\end{split}
\end{equation}
Therefore for $\lambda=1$, we have
\begin{equation}\label{transf3}
\sigma^2_{\lbrace\lambda=1\rbrace}(u_i) =  \Sigma^{2*}_{ii}
\end{equation}
Combining the results in \eqref{transf1} and \eqref{transf3}, we have
\begin{equation}\label{transf4}
\begin{split}
\sigma^{2}_{\lambda}(u_i) = \lambda^{-1} \Sigma^{2*}_{ii},\hspace{3mm}
\sigma^{2}_{\lambda}(u_i) = \lambda^{-1} \sigma^{2}_{\lbrace\lambda=1\rbrace}(u_i)
\end{split}
\end{equation}
This means that for any fixed precision $\lambda$, the marginal standard deviation of the components of a Gaussian vector $\textbf{u}$ can be expressed as a function of $\lambda$ \cite{sorbye2014scaling} by
\begin{equation}\label{transf5}
\sigma_\lambda(u_i) = \dfrac{ \sigma_{\lbrace\lambda=1\rbrace}(u_i)}{\sqrt{\lambda}},\quad i=1,...,n
\end{equation}
For a given IGMRF $\mathbf{u}$ with random precision $\lambda$, we can calculate a reference standard deviation for fixed $\lambda$ = 1, and then approximate the marginal standard deviation for each component of $\mathbf{u}$ \cite{sorbye2014scaling} by\\
\begin{equation}\label{transf6}
\sigma_\lambda(u_i) \approx \dfrac{ \sigma_{ref}(\textbf{u})}{\sqrt{\lambda}} ,\quad i = 1,...,n
\end{equation}
The reference standard deviation is calculated using the geometric mean, an appropriate measure for a set of positive numbers \cite{sorbye2014scaling}. The reference standard deviation for $\mathbf{u}$ in the one-dimensional case is then
\begin{equation}\label{sref1D}
\sigma_{ref}(\mathbf{u}) = \exp\left(\frac{1}{n}\sum_{i=1}^{n}\log\sigma_{\{\lambda=1\}}(u_i)\right)=
 \exp\left(\frac{1}{n}\sum_{i=1}^{n}\frac{1}{2}\log\Sigma_{ii}^*\right)
\end{equation}
where the values $\Sigma_{ii}^*$ denote the diagonal elements of the inverse matrix $\boldsymbol\Sigma^* = \mathbf{Q}^{-1}$ calculated for $\lambda=1$. Specifically, this is calculated as $\textbf{Q}^{-}=\boldsymbol\Gamma^{T}\Lambda^{-}\boldsymbol\Gamma$, where $\boldsymbol\Gamma$ are the eigenvectors and $\boldsymbol\Lambda$ the eigenvalues of $\textbf{Q}$ when $\lambda =1$.
Since $\textbf{Q}$ and $\boldsymbol\Sigma^*$ are both $n\times n$ dimensional for one-dimensional IGMRFs, these $n$ diagonal values are used to calculate the geometric mean. 

We have extended the calculation of $\sigma_{ref}(\textbf{u})$ to two-dimensional second-order IGMRFs. The precision matrix is now $(n_1\times n_2)\times (n_1\times n_2)$ dimensional, where $n_1$ and $n_2$ are the total number of nodes for the first and second variables respectively. The scaling is no longer for $n_1$ or $n_2$ values, but their product $n_1\times n_2$, with a special case when $n=n_1=n_2$
\begin{align}\label{sref2D2}
\sigma_{ref}(\textbf{u}) = \exp\left(\frac{1}{n^2}\sum_{i=1}^{n^2}\log\sigma_{\{\lambda=1\}}(u_i)\right)=
\exp\left(\frac{1}{n^2}\sum_{i=1}^{n^2}\frac{1}{2}\log\boldsymbol\Sigma_{ii}^*\right)
\end{align}
Again, $\boldsymbol\Sigma_{ii}^*$ denotes the diagonal elements of the inverse matrix $(\boldsymbol\Sigma^*)^{1/2} = (\mathbf{Q}^{-1})^{1/2}$, while $\boldsymbol\Sigma^*=\textbf{Q}^{-}=\boldsymbol\Gamma^{T}\Lambda^{-}\boldsymbol\Gamma$ for $\lambda=1$. The precision matrix $\textbf{Q}$, and therefore $\boldsymbol\Sigma^*$, is $n^2 \times n^2$ dimensional, hence there are now $n^2$ elements in the diagonal. 

A further consideration is that for any IGMRF, we must take into account linear restrictions when calculating reference standard deviations so that the latter are finite. Specifically, in the one-dimensional case, for the first-order IGMRF we need to set the last eigenvalue to infinity, and for the second-order IGMRF we set the last two eigenvalues to infinity; for the two-dimensional second-order IGMRF, we must set the three last eigenvalues to infinity \cite{rue2005gaussian}. 

\begin{figure}
	\centering
	\includegraphics[scale=0.6,trim=0 0.5cm 0 2cm]{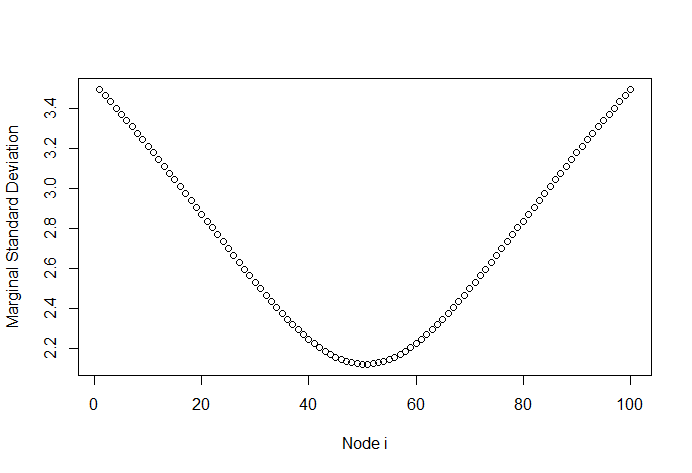}
	\includegraphics[scale=0.6,trim=0 0.5cm 0 0cm]{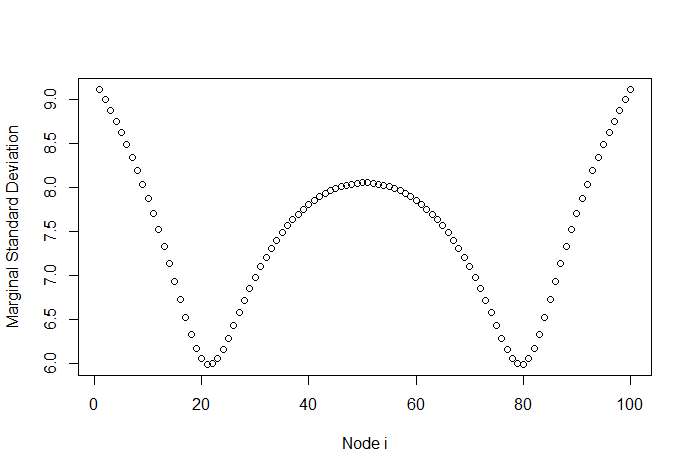}
	\includegraphics[scale=0.7]{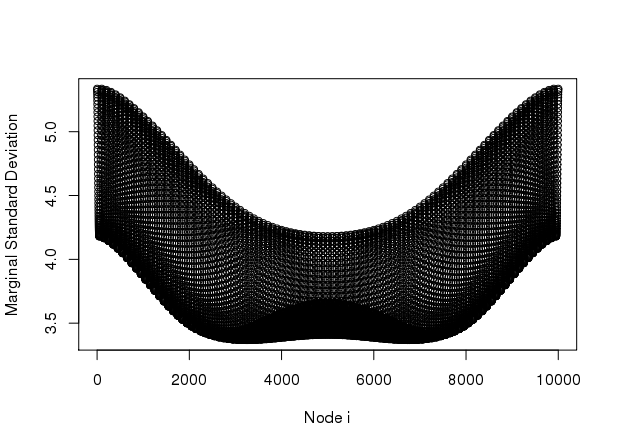}
    \caption{\label{fig:sref}Marginal standard deviations of one-dimensional first-order and second-order, and two-dimensional second-order IGMRFs, calculated using fixed precision $lambda=1$.}
\end{figure}

We see that the new value of the standard deviation in \eqref{b_new} depends on $\sigma_{ref}^{2}$, which captures the precision matrix for a specific type of IGMRF. It is then only necessary to recalculate the standard deviation parameter, $b$, to account for the different shapes and sizes of the graph for a specific IGMRF \cite{sorbye2014scaling}. This can be done for the three types of IGMRF we different specifications of the three IGMRFs lead to shapes and levels of these curves that are quite different, and to substantial variations in their reference standard deviations. It has been shown previously \cite{sorbye2014scaling} that the reference standard deviations in one dimension are $\sigma_{ref}(\textbf{u}) = 3.89$ and $\sigma_{ref}(\textbf{u}) = 41.39$ for the first-order and second-order cases, respectively; applying the result in \eqref{sref2D2}, we find that $\sigma_{ref}(\textbf{u}) = 7.24$ for the two-dimensional second-order case. This means that for a particular hyperprior, larger variances would be allowed for the one-dimensional second-order IGMRF than its two-dimensional equivalent, and both would have larger variances than the one-dimensional first-order case. Equivalently, to allow the same variance, we need to impose an upper limit on the marginal standard deviation
\begin{align}\label{upper_limit1}
Pr(\sigma(u_i)>U)\approx Pr\left(\frac{\lambda}{\sigma^2_{ref}(\mathbf{u})}<\frac{1}{U^2}\right) =\alpha
\end{align}
where $\alpha$ is a fixed small probability \cite{sorbye2014scaling}. By assigning a hyperprior to $\lambda(\sigma^2_{ref}(\mathbf{u}))^{-1}$, the interpretation of the hyperprior remains the same for the different models.

\begin{figure}
	\centering
	\includegraphics[scale=0.7,trim=0 0 2.8cm 0,clip]{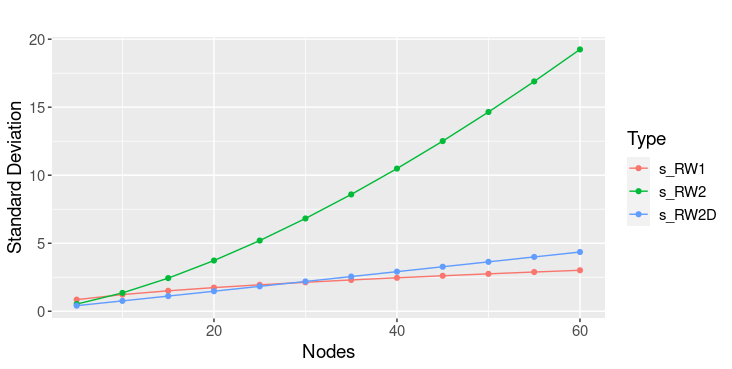}
	\caption{\label{fig:compar}Reference marginal standard deviations for one-dimensional first-order (red), one-dimensional second-order (green), and two-dimensional second-order (blue) IGMRFs with varying numbers of nodes.} 
\end{figure}

These results complement others \cite{lindgren2008second, lindgren2011explicit}, where $k$ equally sized subintervals are created between original nodes $u_1, u_2, \ldots$, to give equidistant nodes 
$u_1^{'},u_2^{'},...,u_{k+1}^{'}$. In the first-order one-dimensional case, the precision using the new nodes is $(k\lambda)^{-1}$, for the second-order equivalent, the precision using the new nodes is $(k^3\lambda)^{-1}$, and finally, for the second-order two-dimensional IGMRF, the precision using the new nodes is $(k^2\lambda)^{-1}$. As shown in Figure \ref{fig:compar}, as the number of nodes increases, the effect of these differences becomes more pronounced.

\subsection{Specifications using Gaussian hyperpriors}
Applying a Gaussian hyperprior, the upper limit expressed in probabilistic form in \eqref{upper_limit1} is 
\begin{align}\label{upper_limit2}
U =\left(\dfrac{b\sigma^2_{ref}(\mathbf{u})}{\Phi^{-1}(\alpha, \mu, 1)} \right)^ {1/2}
\end{align} 
where $\Phi^{-1}(\cdot)$ denotes the quantiles of the Gaussian distribution \cite{sorbye2014scaling}. For a given value of $\alpha$, we can then interpret the mean and standard deviation  parameters, $\mu$ and $b$, in terms of this upper limit. 

To recalculate hyperpriors for different IGMRFs, we can use the same mean parameter $\mu$ for each model and calculate a new standard deviation parameter. By using the upper limit provided in \eqref{upper_limit2}, the new standard deviation parameter is expressed as
\begin{equation}\label{b_new}
 b_{new} = \frac{U^2\Phi^{-1}(\alpha,\mu, 1)} {\sigma^2_{ref}(\textbf{u})}
\end{equation}  
We see that the new value of the standard deviation in \eqref{b_new} depends on $\sigma_{ref}^{2}$, which captures the precision matrix for a specific type of IGMRF. It is then only necessary to recalculate the standard deviation parameter, $b$, to account for the different shapes and sizes of the graph for a specific IGMRF \cite{sorbye2014scaling}. This can be done for the three types of IGMRF considered using
\begin{align}
b_{rw2} = b_{rw1}\times \frac{\sigma^2_{ref}(\mathbf{u}_{rw1})}{\sigma^2_{ref}(\mathbf{u}_{rw2})},\hspace{3mm}b_{rw2D} = b_{rw2}\times \frac{\sigma^2_{ref}(\mathbf{u}_{rw2})}{\sigma^2_{ref}(\mathbf{u}_{rw2D})},\hspace{3mm}b_{rw2D} = b_{rw1}\times \frac{\sigma^2_{ref}(\mathbf{u}_{rw1})}{\sigma^2_{ref}(\mathbf{u}_{rw2D})}
\end{align}
Here, \textit{rw1} and \textit{rw2} refer to the one- and two-dimensional first-order IGMRFs \cite{sorbye2014scaling},  and \textit{rw2D} to the two-dimensional second-order IGMRF. 

\subsection{Types of two-dimensional second order IGMRFs}
We can also compare IGMRFs with fixed order and dimensionality, but different numbers of nodes and boundary conditions. We do so for two-dimensional second-order IGMRFs with four structure matrices: Torus 1 and Torus 2 \cite{rue2005gaussian, thon2012bayesian}, and Bound 1  \cite{yue2010nonstationary} and Bound 2 \cite{terzopoulos1988computation}. Torus 1 has a structure matrix defined on a torus, while Torus 2 has a similar structure matrix but with boundaries at its four corners, $u_{1,1},u_{n_1,1},u_{1,n_2},u_{n_1,n_2}$.
Bound 1 and Bound 2 have boundary effects and induce the same neighbours in the structure matrix for each node, but give different weightings to these neighbours.

In Table \ref{tab:tab1}, we see that Torus 2 consistently has the lowest reference standard deviation, with the changes in each IGMRF being similar proportionally when the number of nodes is increased. Bound 2 has the largest reference standard deviation, followed by Bound 1, which we used in our two-dimensional model of blood pressure \cite{spyropoulou2022}. These findings show that it is clearly necessary to scale the hyperparameter each time the precision matrix or number of nodes is changed, especially when boundary conditions are introduced. 
\begin{table}[h]
	\begin{center}
		\caption{Reference standard deviations $\sigma_{ref}$ for second-order two-dimensional IGMRFs.}
		\label{tab:tab1}
		\begin{tabular}{r r r r r}
		\hline
		Nodes& Torus 1 & Torus 2 & Bound 1  & Bound 2 \\
			\hline
			11&0.58 & 0.10 & 0.83& 1.10\\
			20&1.02 & 0.17 & 1.47& 1.96\\
			40&2.01 & 0.33 & 2.91& 3.87\\
	        100&5.00 &0.83 & 7.24&9.64\\	
	        \hline
	   \end{tabular}
	\end{center}
\end{table}

\vspace{-10mm}
\section{Blood pressure data application}
We compare hyperprior scaling for one- and two-dimensional second-order IGMRFs using blood pressure data \cite{spyropoulou2022}. The scaling varies both by dimensionality and number of nodes, which in our data corresponds to the number of years considered. The hyperpriors must be set for each of the four precision parameters, $\lambda_c$, $\lambda_r$, $\lambda_s$ and $\lambda_g$, that are used at different levels of the hierarchical model: countries, nested in regions, super-regions and the globe. 

When using 40 years of data we find the following values
\begin{align}\label{sigmas}
\sigma_{ref}(\textbf{u}_{rw2})  = 10.486 \qquad
\sigma_{ref}(\textbf{u}_{rw2D}) = 2.91
\end{align}
The distribution under consideration for country-level precision parameters is
$\lambda_c \sim \mathcal{N}(\mu,b)$,
with $\mu$ and $b$  parameters that we assigned the values
\begin{align}\label{known}
\mu = 7, \qquad  
b = 2, \qquad \alpha = 0.001
\end{align}
Here, $b$ is the adjusted parameter to which we need to apply the correct scaling. 
From \eqref{sigmas} and \eqref{known}, the upper levels for the one-dimensional and two-dimensional second-order IGMRFs are 
\begin{equation} \label{U_news}
	\begin{split}
U_{rw2} &= \left(\frac{ b \sigma^2_{ref}(\textbf{u}_{rw2})}{\Phi^{-1}(\alpha,\mu,1)}\right)^{1/2}  = 7.5\\
U_{rw2D} &= \left(\frac{b \sigma^2_{ref}(\textbf{u}_{rw2D})}{\Phi^{-1}(\alpha,\mu,1)}\right)^{1/2}= 2.08
\end{split}
\end{equation}
We do not have negative values in the Gaussian quantiles, so to reproduce earlier results \cite{sorbye2011simultaneous} using a Gaussian rather than a Gamma distribution, we need to proceed as if we have truncation below at zero. By taking the median, we have
\begin{align}
\text{median}(U_{rw2},U_{rw2D}) = U = 4.79
\end{align}
The new standard deviation parameters for the hyperpriors are:
\begin{equation} \label{b_news}
	\begin{split}
b_{rw2} &= \frac{U^2 \Phi^{-1}(\alpha,\mu,1)} {\sigma^2_{ref}(\textbf{u}_{rw2})} = 0.81\\
b_{rw2D} &= \frac{U^2\Phi^{-1}(\alpha,\mu, 1)}{\sigma^2_{ref}(\textbf{u}_{rw2D})} = 10.59
\end{split}
\end{equation}
Alternatively knowing $b_{rw2}=0.81$,
\begin{align*}
	b_{rw2D} &= b_{rw2} \frac{\sigma^2_{ref}(\textbf{u}_{rw2})}{\sigma^2_{ref}(\textbf{u}_{rw2D})} = 10.59
\end{align*}

As in Figure \ref{fig:compar}, Table \ref{tab:tab2} shows that the one-dimensional second-order IGMRF has the largest variation as the number of nodes is increased. 

We also observe different patterns as the adjusted parameter, $b$, is varied. In earlier work \cite{danaei2011national}, the standard deviation of the one-dimensional second-order IGMRF, $b_{rw2}$, was set to 3. In the case of five nodes, scaling makes this equivalent to 4.96 for the two-dimensional case, $b_{rw2D}$, but here the adjusted parameter, $b$, is equal to 3, as shown in Table \ref{tab:tab3}. We also see variations in the tuning of $b_{rw1}$, $b_{rw2}$ and $b_{rw2D}$ as the number of nodes changes, and in particular cases, each of them coincides with the adjusted parameter $b$. For example, we see that the adjusted parameter, $b$ is equal to $b_{rw1}$ when the numbers of nodes are 11 and 20; for five nodes, the adjusted parameter is equal to $b_{rw2}$; while for 40 nodes, $b_{rw2D}$ is equal to the adjusted parameter.

\begin{table}[ht]
\begin{center}
\caption{Reference standard deviations, $\sigma_{ref}$, for models with 11 and 20 nodes, for one-dimensional first-order and second-order IGMRFs, and two-dimensional second-order IGMRFs. }
		\label{tab:tab2}
		\begin{tabular}{r r r r}
		\hline
		Nodes& $\sigma_{ref}(\textbf{u}_{rw1})$&$\sigma_{ref}(\textbf{u}_{rw2})$&$\sigma_{ref}(\textbf{u}_{rw2D})$\\
		\hline
	11&1.28&1.54&0.83\\
	20&1.74&3.73&1.47\\
	\hline
		\end{tabular}
	\end{center}
\end{table}

\begin{table}[ht]
	\begin{center}
	\caption{Scaling the standard deviation parameters, $b_{rw1}$, $b_{rw2}$ and $b_{rw2D}$, as the adjusted parameter $b$ and number of nodes are varied.}
		\label{tab:tab3}
		\begin{tabular}{r r r r r r r r r r}
		\hline
		\multirow{2}{10.5mm}{Nodes}&\multicolumn{3}{c}{$b$=1}&\multicolumn{3}{c}{$b$=2}&\multicolumn{3}{c}{$b$=3}\\
		\cline{2-10}
 & $b_{rw1}$ & $b_{rw2}$ & $b_{rw2D}$ & $b_{rw1}$ & $b_{rw2}$ & $b_{rw2D}$ & $b_{rw1}$ & $b_{rw2}$ & $b_{rw2D}$\\
		\hline
		5& 0.38 &1.00&1.65&0.76&2.00&3.30&1.14&3.00 &4.96\\
		11&1.00&0.69& 2.39&2.00&1.39&4.78&3.00&2.08&7.17\\
		20&1.00&0.22& 1.39&2.00 &0.43 &2.78&3.00 &0.65 &4.17\\
		40&1.4& 0.08& 1.00&2.80&0.15&2.00&4.2 &0.23 &3.00\\
		\hline
		\end{tabular}
	\end{center}
\end{table}

\begin{table}[ht]
\begin{center}
\caption{Scaling the standard deviation parameters $b_{rw2}$ and $b_{rw2D}$ for one- and two-dimensional IGMRFs with 11 nodes for $\lambda_c$, $\lambda_r$, $\lambda_s$ and $\lambda_g$, as $b$ is varied.}
		\label{tab:tab4}
		\begin{tabular}{l r r r}
		\hline
		$\lambda$&$b$& $b_{rw2}$&$b_{rw2D}$\\
		\hline
	$\lambda_c$&0.9&0.53&1.83\\
	$\lambda_r$&1.2&0.71&2.44\\
	$\lambda_s$&1.59&0.94&3.24\\
	$\lambda_g$&3.55&2.10&7.23\\
	\hline
		\end{tabular}
	\end{center}
\end{table}

Table \ref{tab:tab4} shows the scaling applied to the standard deviations of $\lambda_c$, $\lambda_r$, $\lambda_s$ and $\lambda_g$. For example, for an adjusted parameter $b=0.9$ and 11 nodes, applying equations \eqref{known} to \eqref{b_news} to scale $\lambda_c$ gives 0.53 for the one-dimensional second-order model, and 1.83 for the two-dimensional equivalent.  We also see differences in results between Tables \ref{tab:tab3} and \ref{tab:tab4}. For the results in Table \ref{tab:tab3}, the upper level is defined by the median of three values, $b_{rw1}$, $b_{rw2}$ and $b_{rw2D}$, whereas the values in Table \ref{tab:tab4} are based on the median of two values, $b_{rw2}$ and $b_{rw2D}$. This variation in $U$ then changes the values of $b_{rw2}$ and $b_{rw2d}$, with these differences becoming more apparent as the number of nodes increases. These results allowed us to apply the same degree of smoothness in the two-dimensional second-order case as in the earlier work, scaling $b_{rw2D}$ correctly given the variation in the blood pressure data.

\section{Summary and future work}
We have shown the importance of correct scaling of hyperpriors for precision parameters in IGMRFs. This scaling varies by the dimensionality, order, and size of the IGMRFs, and also depends on the structure of the precision matrices, with substantial variations. We have shown both general results in two dimensions and a specific application to a two-dimensional model of blood pressure data.

Future work could include applying penalised complexity (PC) priors as precision parameters for two-dimensional random effects \cite{simpson2017penalising}. The precision parameter corresponds to a second-order IGMRF, $\mathbf{u}\sim \mathcal{N}(0,\lambda^{-1}\mathbf{P}^{-1})$. However, a model can have two types of random effects, constructed and unconstructed. They have dependent precision parameters, and so a joint bivariate or multivariate distribution should express this dependence. For the two-dimensional second-order IGMRF, the precision parameter is univariate but we could investigate the use of PC priors, which have the property that no further scaling is required as the number of nodes is varied.
\vspace{4mm}\\
\noindent\textbf{Acknowledgements} Specialist and High Performance Computing systems were provided by Information Services at the University of Kent.
\bibliographystyle{apacite}
\bibliography{bibl}
\end{document}